# Experimental Performances Analysis of Load Balancing Algorithms in IEEE 802.11

HAMDI Salah[1]

Computer Sciences Department

ISSAT of Sousse

Sousse, Tunisia

[1]hamdisalah@yahoo.fr

SOUDANI Adel[2], TOURKI Rached[3]

Laboratory of Electronic and Microelectronic

Sciences Faculty of Monastir

Monastir, Tunisia

[2]adel.soudani@issatso.rnu.tn, [3]rached.tourki@fsm.rnu.tn

*Abstract—* **In IEEE 802.11, load balancing algorithms (LBA) consider only the associated stations to balance the load of the available access points (APs). However, although the APs are balanced, it causes a bad situation if the AP has a lower signal length (SNR) less than the neighbor APs. So, balance the load and associate one mobile station to an access point without care about the signal to noise ratio (SNR) of the AP cause possibly an unforeseen QoS; such as the bit rate, the end to end delay, the packet loss, … In this way, we study an improvement load balancing algorithm with SNR integration at the selection policy.**

## I. INTRODUCTION

At the time of communication process, one mobile station selects always the near access point who gives the most excellent signal length among those of all available APs. However, the client number per AP increases, so the bit rate per client and the network performance decreases. In the different standards IEEE 802.11, the association decision of a mobile station to an access point is made only thanks to physique consideration without care about load of the APs. In fact, many access points will be more loaded than the other neighbor APs and the quality of service decrease. In this way, many techniques and approaches are proposed to resolve this problem of unbalanced load in IEEE 802.11. Usually, the approaches propose load balancing algorithms (LBA) to equilibrate traffic between the different available Wi-Fi nodes.

In this paper, we show an experimental analysis of QoS and of load balancing algorithm in IEEE 802.11. The paper is organized as follows: in section 2, we outline the problem of unbalanced load. In section 3, we show many different approaches focalized about this problem. In section 4, we address the limit of LBA. In section 5, we have used an experimental platform (camera IP transmitting video MPEG-4, APs, mobiles stations …) to apply the algorithm and to do many different experiences in IEEE 802.11environment. We have applied LBA proposed in [4, 5, 8] and we have analyzed his efficiency. Section 6 presents our contribution to improve LBA. Finally, section 7 concludes this work.

## II. PROBLEM OF UNBALANCED LOAD IN IEEE 802.11

When load balancing word is used in IEEE 802.11, load means the number of the active process per access point and a load balancing mechanism attempt to make the same number of active process per cell [10]. The standard IEEE 802.11 does not specify an automatic load distribution mechanism. In the hot spots who dispose of many distributed access points, one mobile station selects always an AP who gives the most excellent signal to noise ratio (SNR). The users search the near AP without care about the traffic state of the selected AP. In fact, this phenomenon causes a problem to wireless LAN who is not dimensioned and many APs are managing several mobiles more than the available neighbor APs. In this way, upload an access point more than another AP cause an unbalanced load problem. Figure 1 show that one mobile station who is moving between several APs do not have QoS criterion to help it to choice one AP and not choice another.

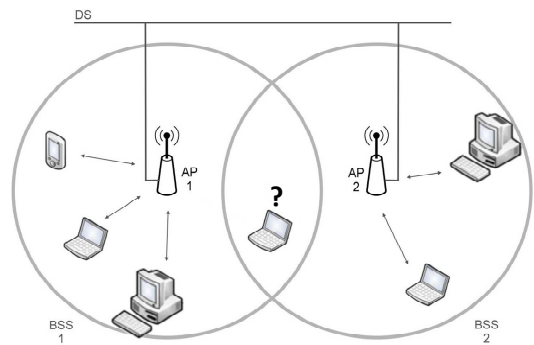

Fig. 1. Unbalanced load problem

Load balanced algorithm is applied in the intersection zones of the different APs and one mobile station is attached from an access point to another.

## III. PREVIOUS LOAD BALANCING APPROACHES

I. Papanikos and all in [2] indicate that load balancing policy is necessary to distribute the mobiles stations between the different access points. They have proposed one load balancing procedure to attach one mobile station to an AP and balance the user's number per AP.

S. Sheu and all in [1] proposed Dynamic Load Balance Algorithm (DLBA) to distribute the WLAN users according to the client number per AP.

[9] Proposed an algorithm to decrease congestion and balance the user's traffics in IEEE 802.11. Each access point has one different AP channel to avoid congestion event and signal interference problems. The algorithm found the Most Congested Access Point (MCAP), analyses the users association and decrease the congestion of MCAP. A user will not be connected to an AP with the result that the Signal to Interference Ratio (SIR) is absolute positive and the signal power is more than a fixed threshold.

[7] Presented a Cell Breathing Technique to balance the APs load and improve QoS of the real time applications. It reduces the signal length of the congested APs, so reduce the AP's impact and user's number per congested AP. On the other hand, it increase the signal length and the Impact of the under loaded APs. So reattach the disconnected stations to the under loaded access point.

V. Aleo and all in [4, 5] proposed a load balancing algorithm in hot spots. The algorithm works with care about the bit rate per each AP. It does not think about the client's number per AP because the user's traffic is changeable. So, there are not a serious correlation between the client's number and their traffics. Over loaded access points are not authorized to associate new coming stations.

### IV. LOAD BALANCING ALGORITHM LIMITS

Wireless link characteristics are not constant and vary over time and place [6]. Load balancing algorithms consider only the associated stations to balance the load of the all APs. However, although the APs are balanced, it cause a bad situation if the AP's associated stations is having low signal length (SNR) less than the neighbor APs. Possibly, it will suffer the AP channel and increase the number of loss packets. If IEEE 802.11 is not dimensioned correctly and the APs are distributed wrongly, so it's impossible to apply load balancing algorithm and improve the QoS [3]. Moreover, associate a mobile station to an access point without consideration of the signal length (SNR) received from the AP, cause possibly an unforeseen QoS; such as the bit rate, the end to end delay, the packet loss, … On the other hand, an under loaded access point but having low Signal to Noise Ratio (SNR) cannot improve QoS. In fact, before apply LBA and change one mobile station from an AP to another, it's very important to think about noise, signals interference, distance and geographic distribution of the available APs and so their signals levels. An access point having low SNR must not consider at the time of LBA execution. In this way, we search to show this contribution experimentally. We use an experimental platform with camera IP and many APs to analyze QoS of MPEG-4 transmission via IEEE 802.11 and measure many different parameters.

### V. EXPERIMENTATIONS & RESULTS

#### A. Exp1 : insuffisance of SNR

This first experimentation has an object to analyze the video MPEG-4 quality and measure many different parameters of the QoS. We show the variation of bit rate, end to end delay, jitter, loss … according to SNR and load of the APs.

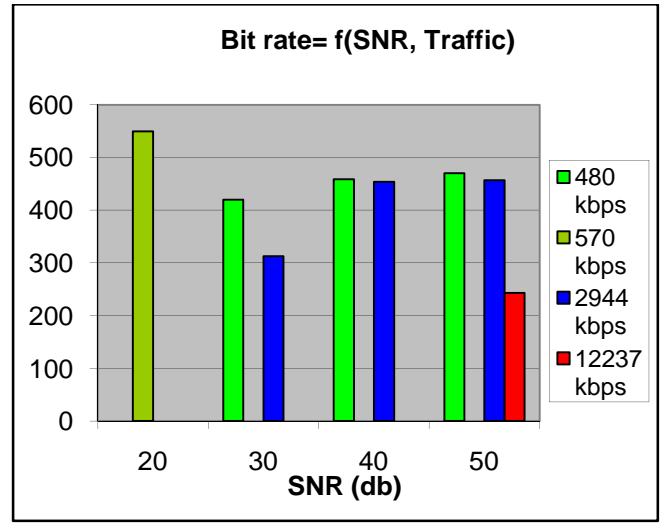

Fig. 2. Bit rate variation

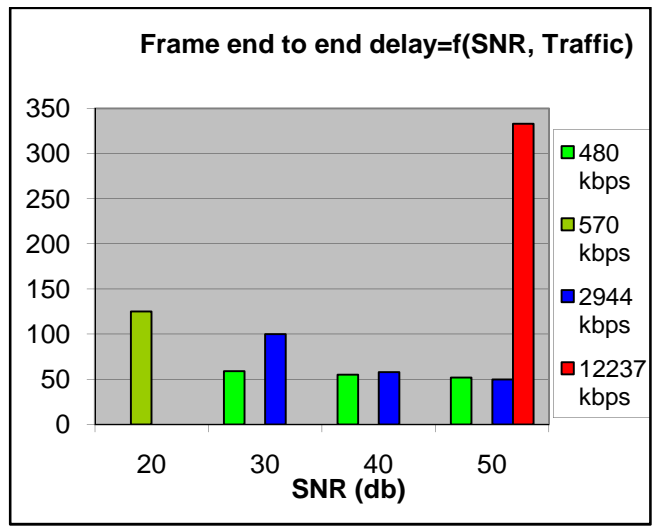

Fig. 3. Frame end to end delay variation

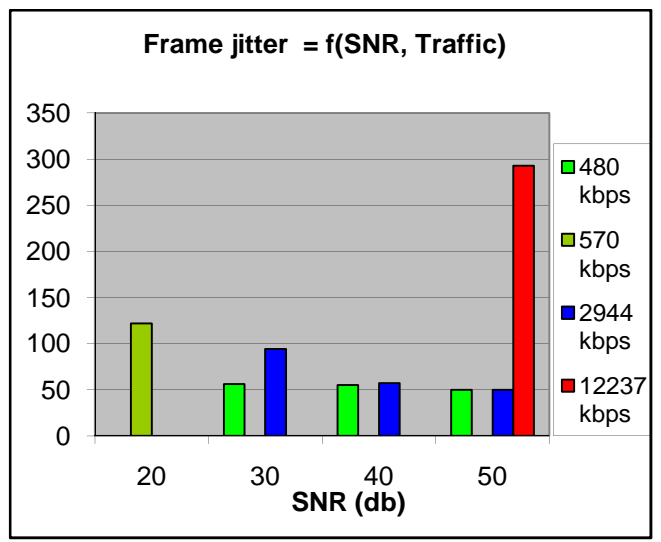

Fig. 4. Frame jitter variation

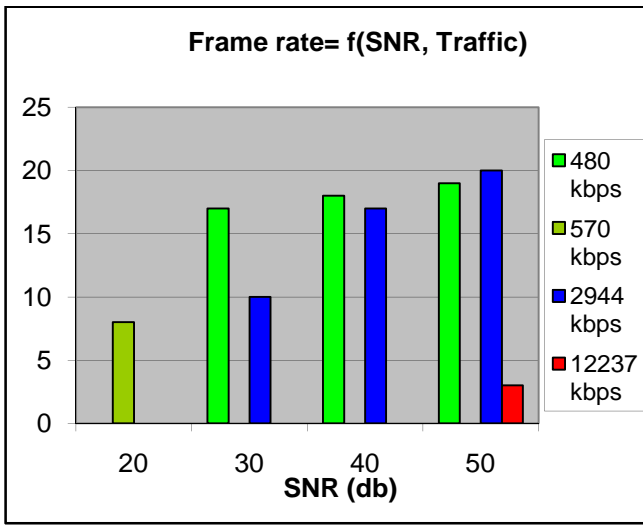

Fig. 5. Frame rate variation

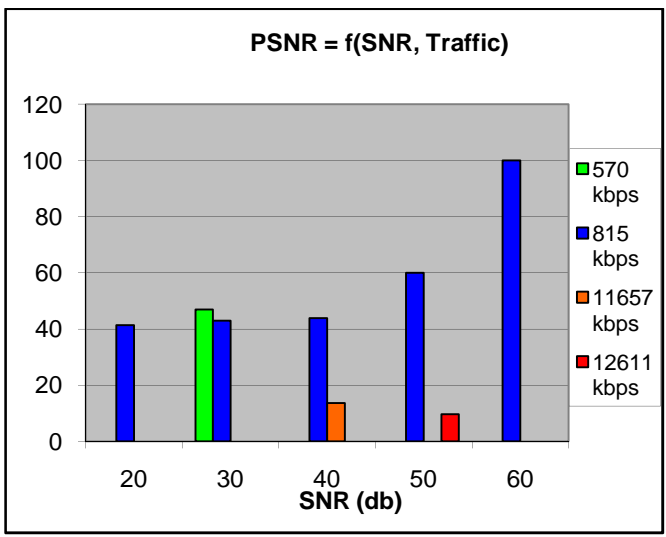

Fig. 6. PSNR variation

Figure 2 show the bit rate variation according to SNR and load. The bit rate varies proportionally to SNR. However, if AP's load increase so the bit rate decrease although SNR is strong again ; for example, when the traffic is equal to 12237 kbps and SNR = 50 db, the bit rate (243 kbps) is less than the value (550 kbps) measured when the traffic = 570 kbps although SNR is weak.

Figure 3 demonstrates that the delay does not vary proportionally to SNR. Although, SNR is strong (50 db) and traffic = 12237 kbps, the frame delay is more (333 msec) than the value measured (59 msec) when the traffic = 480 kbps and SNR = 30 db.

As figure 4 shows, frame jitter does not vary proportionally to Signal to Noise Ratio. However, if an access point is uploaded so the jitter increases although SNR value is good. We have measured a very bad jitter (293 msec) when the traffic =12237 kbps and SNR=50 db. On the other hand, we have measured good jitter (56 msec) when the traffic = 480 kbps and SNR was medium (30 db).

In figure 5, we have measured the frame number received per second according to (SNR) and the load of AP. If SNR is good so the frame rate increases. But the rate decrease (3 fps) when the traffic = 12237 kbps although SNR is good (50 db) because AP is uploaded. However, the rate was good (17 fps) when the traffic = 480 kbps although SNR was medium (30 db).

Figure 6 demonstrates that video quality and so PSNR increase according to the signal length. However, the traffic of access point affects the video quality.

Basing on the previous figures and interpretations, QoS parameters are better if Signal to Noise Ratio is strong. However, upload an AP decrease the QoS although SNR is good again. In fact, take care only about SNR and physiques criterion of channel is not sufficient to improve QoS. We show experimentally the importance of the load at the variation of QoS parameters at the time of MPEG-4 transmission. We must considerer AP's load at the time of IEEE 802.11 connection.

*B. Exp2: Performance analyze of LBA*

The object of this second experience is to use again our platform (camera IP transmitting video MPEG-4, APs, mobiles stations …) and apply LBA between two access points that are unbalanced. The standard IEEE 802.11 does not distribute automatically the traffic of APs. Indeed, we apply LBA manually basing on the sum of traffics and we distribute one mobile station to balance the load of the two available access point.

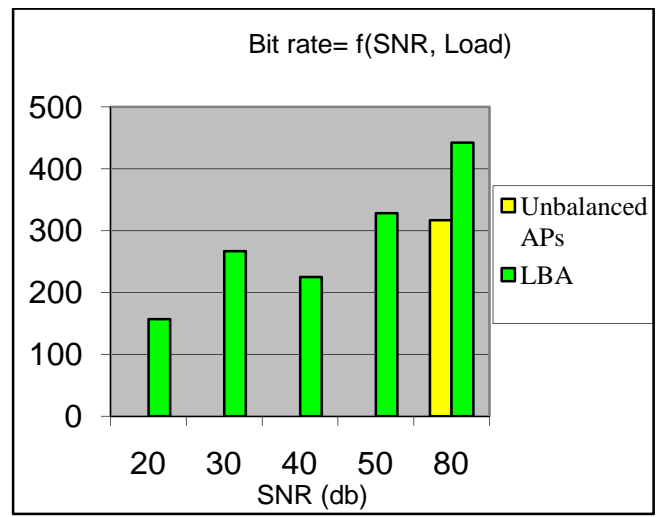

Fig. 7. Bit rate variation

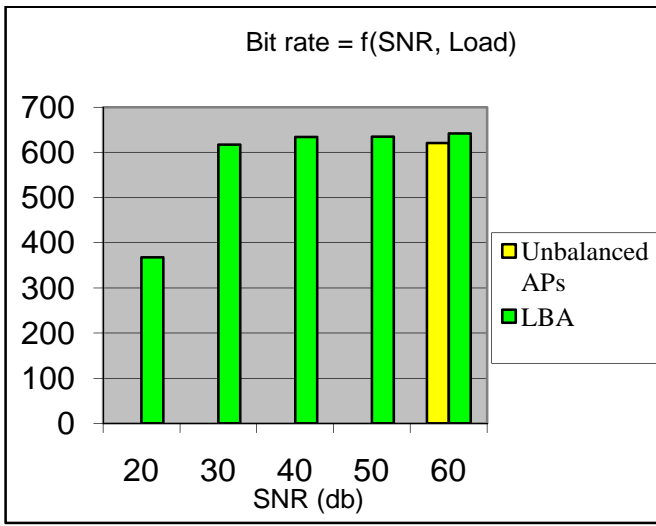

Fig. 8. Bit rate variation

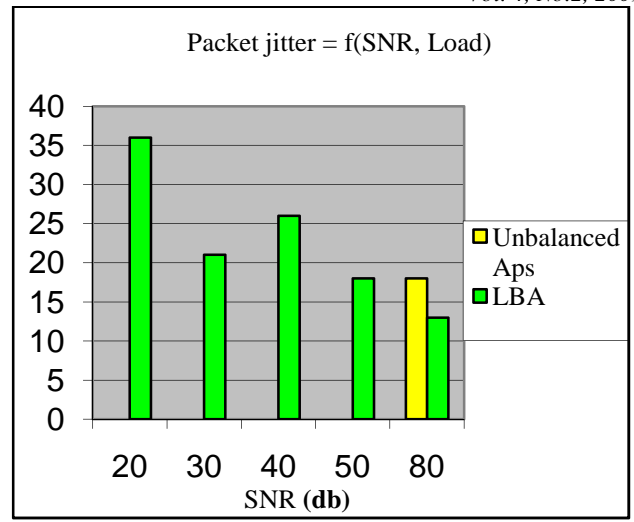

Fig. 10. Packet jitter variation

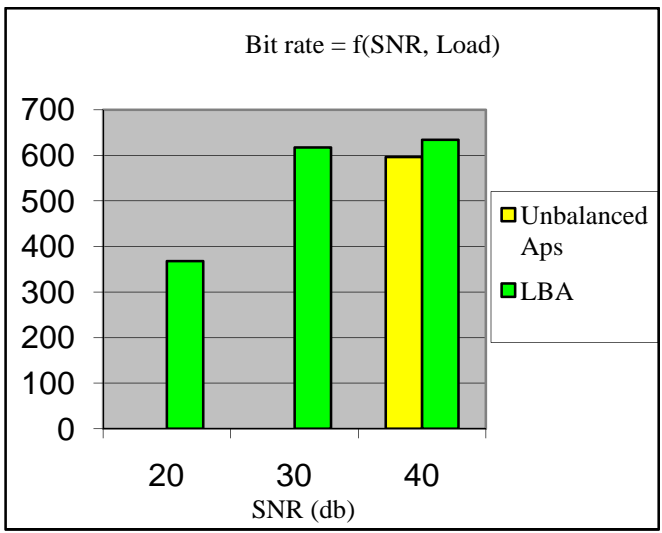

Fig. 9. Bit rate variation

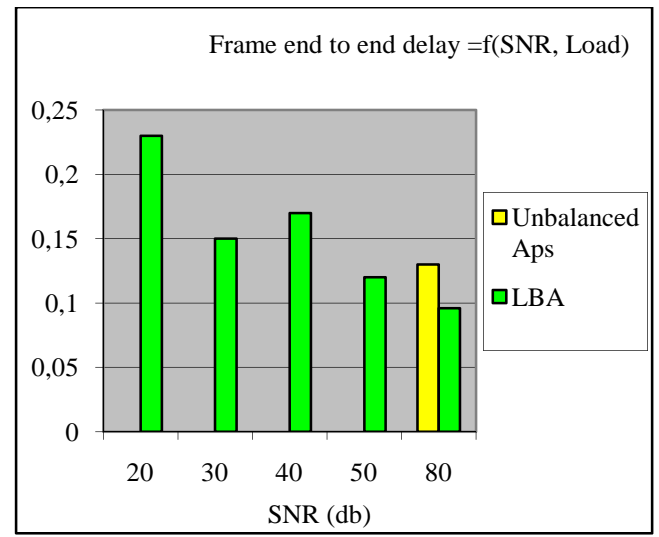

Fig. 11. Frame end to end delay variation

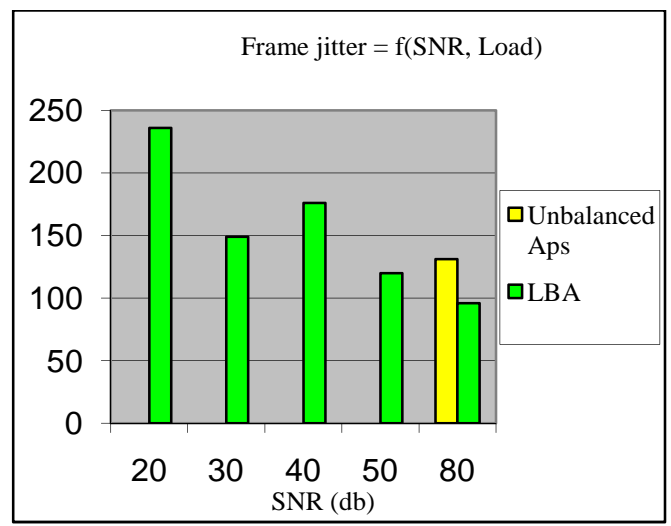

Fig. 12. Frame jitter variation

At the first hand, figures 7, 8 and 9 show the efficiency of load balancing algorithm to enhance the bit rate of users. In figure 7, we have applied LBA at one signal length = 80 db so the rate increase from 317 kbps to 442 kbps.

On the other hand, the bit rate decreases again while SNR is weak. Although APs are balanced, we have measured 225 kbps who is less than the first value measured when the APs were unbalanced (317 kbps). This last note is valid on figures 8 and 9. However, we balance the APs at a signal length =SNR1, so the bit rate increase. But it decrease again at a weak signal length = SNR2 and we have measured bad value who is less than the value calculated at SNR1.

Figures 7, 8 and 9 show a correlation between SNR1 and SNR2; the bit rate decrease while SNR2 = SNR1/2 although APs are balanced again (SNR2 = 40 db if SNR1 = 80 db, SNR2 = 30 db if SNR1 = 60 db, SNR2 = 20 db if SNR1 = 40 db).

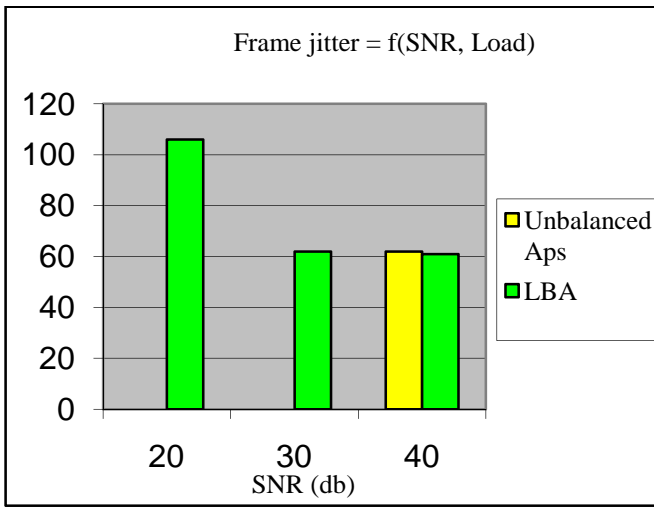

Fig. 13. Frame jitter variation

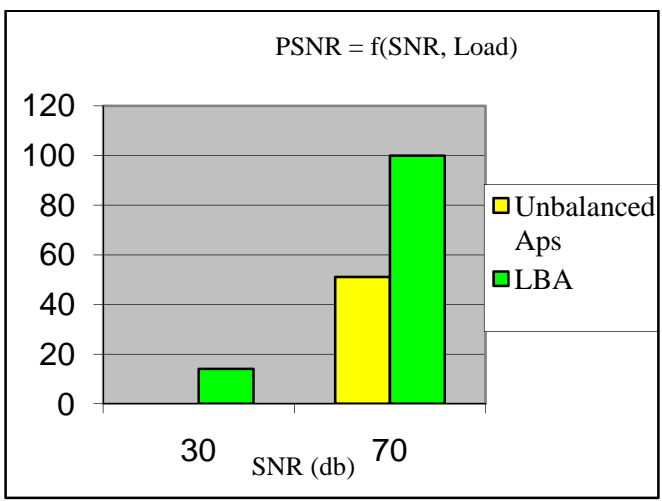

Fig. 14. PSNR variation

In figure 10, we have applied LBA when signal to noise ratio = 80 db. In fact, packet jitter decreases from18ms to 13ms. However, the jitter increase again according to SNR (<=40 db) until 26 ms that is greater than the first value (18 ms) measured when APs are unbalanced.

As figure 11 shows, LBA application decrease considerably frame end to end delay from 0.13 sec to 0.096 sec. Although the access point are balanced, the end to end delay increase until 0.17 sec because SNR became weak (SNR <= 40 db) than the first value (80 db).

In figure 12, LBA application improves the frame jitter. In fact, the jitter decreases from 131 ms until 96 ms. But looking at the value measured in figure 13 (176 ms) when SNR became weak (40 db), frame jitter is bad than the value calculated when SNR is strong (80 db) although APs were unbalanced (131 ms).

Figure 14 demonstrates that LBA enhance video quality of users. Figure 14 show that when SNR = 70 db, LBA application increases PSNR considerably. However, when signal length became medium = 30 db, video quality became bad and so PSNR decrease (14 db) although APs are balanced.

On conclusion, the previous figures were an application of load balancing algorithm to study his efficiency. Basing on the figures, LBA improve QoS and enhance their parameters (bit rate, jitter, end to end delay ...). However, the stations are mobiles and signals lengths vary from one mobile station to another. Indeed, LBA application is not absolutely the best solution to improve the QoS. Quality of service decrease when SNR became weak although the load is distributed correctly. So, we show experimentally the importance of parameter SNR at the time of LBA application. In fact, apply LBA and change one mobile station from an uploaded AP to an under loaded AP do not improve inevitably QoS. The second AP must not have SNR who is less than the half of the first SNR.

## VI. CONTRIBUTION TO LBA ENHANCEMENT

LBA show a limit between selection policy and distribution policy. Selection policy select one mobile station that will be disconnected from an uploaded AP and it will be connected to an under loaded AP. Then, distribution policy checks the load balancing criterion β and distributes the selected mobile. Indeed, selection policy thinks without care about Signal to Noise Ratio (SNR) of the available under loaded and uploaded APs. In fact, an under loaded AP can have bad physique criterion and so weak SNR although it's more available than the author APs. However, although APs are now balanced, QoS decrease when the new under loaded APs or their associated stations are far off than the old uploaded AP. So, it will suffer AP's channel and have a very high probability of packet loss. In this way, we try to improve load balancing algorithm with integration of parameter SNR at the selection policy (figure 15).

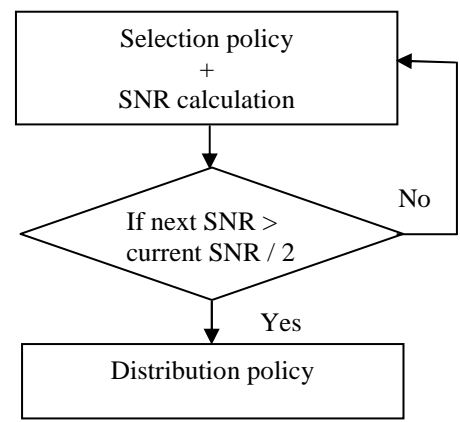

Fig. 15. LBA enhancement

Our contribution to enhance LBA means the next: at the time of LBA application, if LBA decide to disconnect one mobile from an AP and connect it to another, so it's necessary to think with care about the SNR of the new AP. The new signal to noise ratio must not less than the half of the SNR of the old AP although the new AP is under loaded.

## VII. CONCLUSION AND FUTURE WORKS

In this paper, we have proposed a contribution to improve the QoS of video MPEG-4 transmission via IEEE 802.11. In this way, we have used an experimental platform (camera IP, APs, mobiles…) to do two experiences. At the first hand, we have studied QoS parameters (bit rate, jitter, end to end delay…) variation according to SNR and load. On the other hand, we have analyzed the performances of load balancing algorithm in IEEE 802.11. These experiences allow us to found many results: firstly, QoS vary proportionally to signal to noise ratio SNR but load of APs affect QoS in IEEE 802.11. Secondly, LBA works more with taking care about SNR of the available APs. Finally, we have proposed a new approach basing on SNR at the time of LBA execution.

The future works should be focus on the new approaches and primitives that can be introduced to enhance the QoS. We will study the implementation of load balancing algorithm with SNR integration at the selection policy. In this way, we can use a network simulator such as OPNET or NS to simulate and test these strategies.

AUTHORS PROFILE

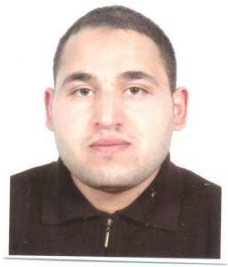

**Salah HAMDI** Received his Teaching and Master degrees in computer sciences from the Upper Institute of the Applied Sciences and Technology (ISSAT) of Sousse, Tunisia. Currently, he is PhD student in National Engineering School of Sfax (ENIS), Tunisia. His current research topics concern the artificial intelligence and the help to decision. He focuses on the design of intelligent software for the decision in cardiology.

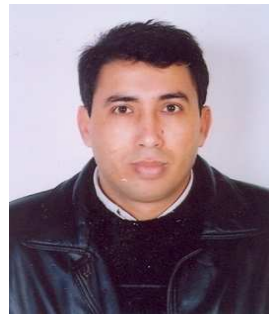

**Adel SOUDANI** received his PhD (2003) in Electronics and also Electrical Engineering respectively from the University of Monastir, Tunisia, and the University of Henri Poincaré Nancy I, France. He is currently an Assistant Professor at the Institute of Applied Sciences and Technology of Sousse. His research activity includes QoS management in real time embedded systems and multimedia applications. He focuses mainly on protocol verification, implementation and performance evaluation for multi-constrained communication systems.

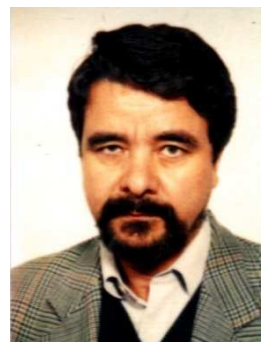

**Rached TOURKI** received the B.S. degree in Physics (Electronics option) from Tunis University, in 1970; the M.S. and the Doctorat de 3eme cycle in Electronics from Institut d'Electronique d'Orsay, Paris-south University in 1971 and 1973 respectively. From 1973 to 1974 he served as microelectronics engineer in Thomson-CSF. He received the Doctorat d'etat in Physics from Nice University in 1979. Since that date, he has been professor in Microelectronics and Microprocessors in the department of physics at the Science Faculty of Monastir. From 1999, he is the Director of the Electronics & Microelectronics Lab.